\documentclass{PoS}

\newcommand{\bk}{\mbox{\boldmath $k$}}

\newcommand{\bPhi}{\mbox{\boldmath $\Phi$}}
\newcommand{\bDelta}{\mbox{\boldmath $\Delta$}}

\def\lsim{\mathrel{\rlap{\lower4pt\hbox{\hskip1pt$\sim$}}
    \raise1pt\hbox{$<$}}}         
\def\gsim{\mathrel{\rlap{\lower4pt\hbox{\hskip1pt$\sim$}}
    \raise1pt\hbox{$>$}}}         
\def\Pom{{\bf I\!P}}

\title{Diffractive production of quark-antiquark pairs}

\ShortTitle{Diffractive production of quark-antiquark pairs}

\author{\speaker{Antoni SZCZUREK}
\thanks{This work was supported by the Polish National Science Centre 
(on the basis of decision No.~DEC-2011/01/B/ST2/04535).}\\
University of Rzesz\'ow, PL-35-959 Rzesz\'ow, Poland, and\\
Institute of Nuclear Physics PAN, PL-31-342 Cracow, Poland\\
E-mail: \email{Antoni.Szczurek@ifj.edu.pl}}

\author{Marta {\L}USZCZAK\\
University of Rzesz\'ow, PL-35-959 Rzesz\'ow, Poland\\
E-mail: \email{luszczak@univ.rzeszow.pl}}

\author{Wolfgang SCH\"AFER\\
Institute of Nuclear Physics PAN, PL-31-342 Cracow, Poland\\
E-mail: \email{Wolfgang.Schafer@ifj.edu.pl}}

\abstract{
We discuss diffractive dissociation of gluons into heavy quark pairs.
The particular mechanism is similar to the diffractive dissociation
of virtual photons into quarks. 
The amplitude for the $g p \to Q \bar Q p$ is derived in the impact
parameter and momentum space. The cross section for single diffractive
$p p \to Q \bar Q p X$ is calculated as a convolution of the elementary
cross section and gluon distribution in the proton.
Integrated cross section and differential distributions in transverse 
momentum and rapidity of the charm and bottom quark and antiquark, etc. 
are calculated for the nominal LHC energy for different unintegrated 
gluon distributions from the literature.
The ratio of the bottom-to-charm cross sections are shown and discussed
as a function of several kinematical variables. 
}

\FullConference{XXI International Workshop on Deep-Inelastic Scattering
  and Related Subject -DIS2013,\\
		22-26 April 2013\\
		Marseilles,France}

\begin{document}

\section{Introduction}

Hard diffractive production is characterized by the production of
massive objects
($W^{\pm}$, $Z^0$, Higgs boson, pairs of heavy quark - heavy antiquark)
or objects with large transverse momenta (jets, dijets) and one (single
diffractive process) or two (central diffractive process) rapidity gaps
between proton(s) and the centrally produced massive system.
The cross section for these processes is often calculated in terms of 
hard matrix elements for a given process and so-called diffractive
parton distributions.
The latter are often calculated, following a suggestion of Ingelman and 
Schlein\cite{IS}, in a purely phenomenological approach in terms of 
parton distributions in the pomeron and a Regge-theory motivated flux 
of Pomerons. 

Diffractive production of heavy quarks was previously discussed 
within the Ingelman-Schlein model in
Refs.\cite{Fritzsch:1985wt,diffractive_open_charm_1,diffractive_open_charm_2,LMS2011}
and proposed as a probe of the hard substructure of the Pomeron.

In this presentation we discuss a specific mechanism for the 
diffractive production of heavy quark -- antiquark pairs in 
proton-proton collisions in a ``microscopic approach'' which does not 
use the assumptions of Regge factorization, and in which the QCD Pomeron
is rather modelled by exchange of a gluon ladder related to the
unintegrated gluon distribution in the proton.

The mechanism we propose is based on the partonic subprocess 
$g p \to Q \bar Q p$ -- the diffractive dissociation of a gluon into 
a heavy quark pair.  
The forward amplitude for the $g p \to Q \bar Q p$ is well behaved 
and is perturbatively calculable without introducing new soft parameters. 


In the usual treatment of hard diffraction, heavy quarks are generated 
from gluons in the Pomeron and a valence-like heavy quark contribution 
is not present. In this sense the mechanism discussed here is
complementary to existing approaches, although eventually we intend that
the Ingelman-Schlein mechanism be superseded also by a microscopic model
for the gluon distribution in the Pomeron. 

\begin{figure}[!h]
\includegraphics[width=6cm]{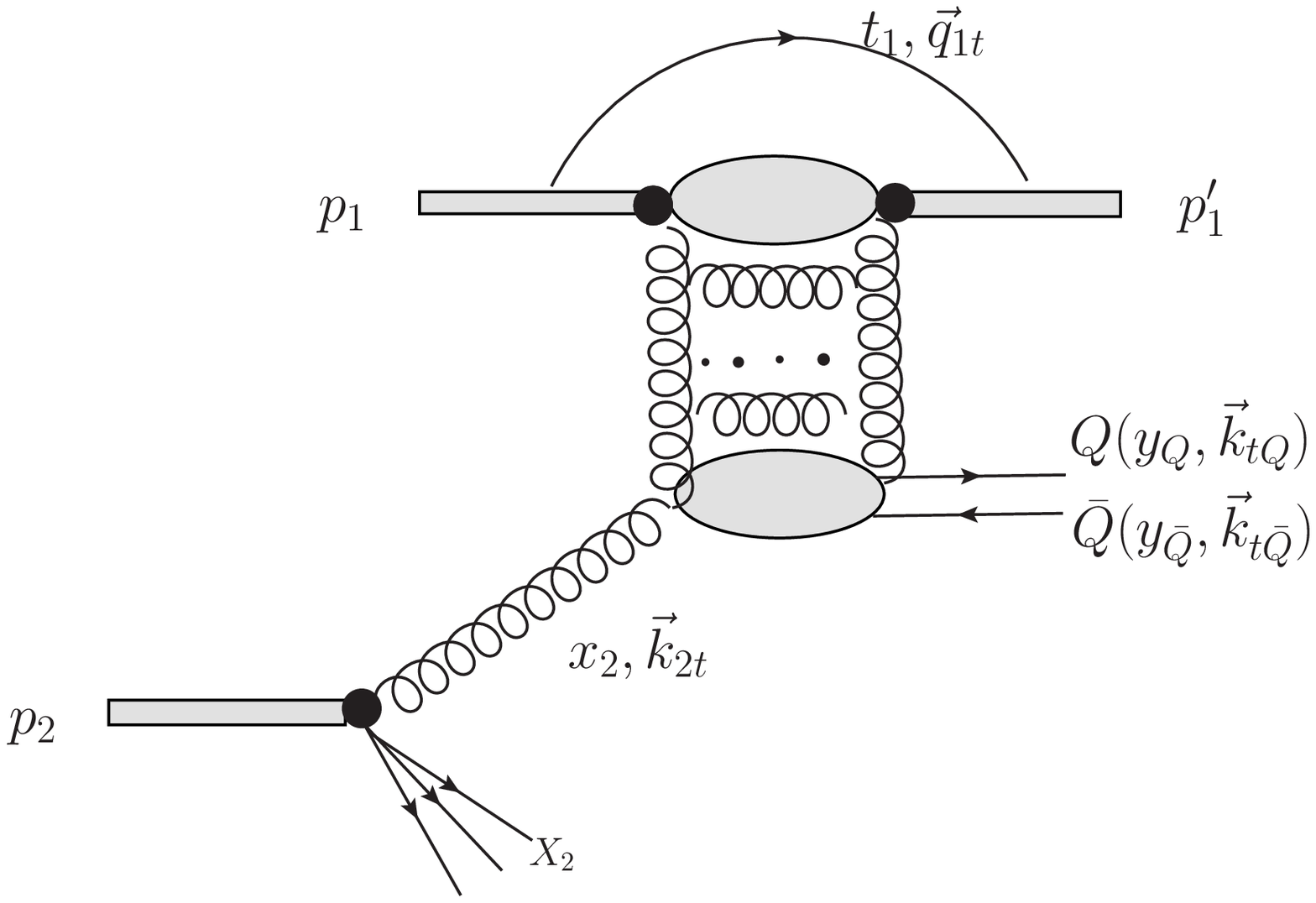}
\includegraphics[width=7cm]{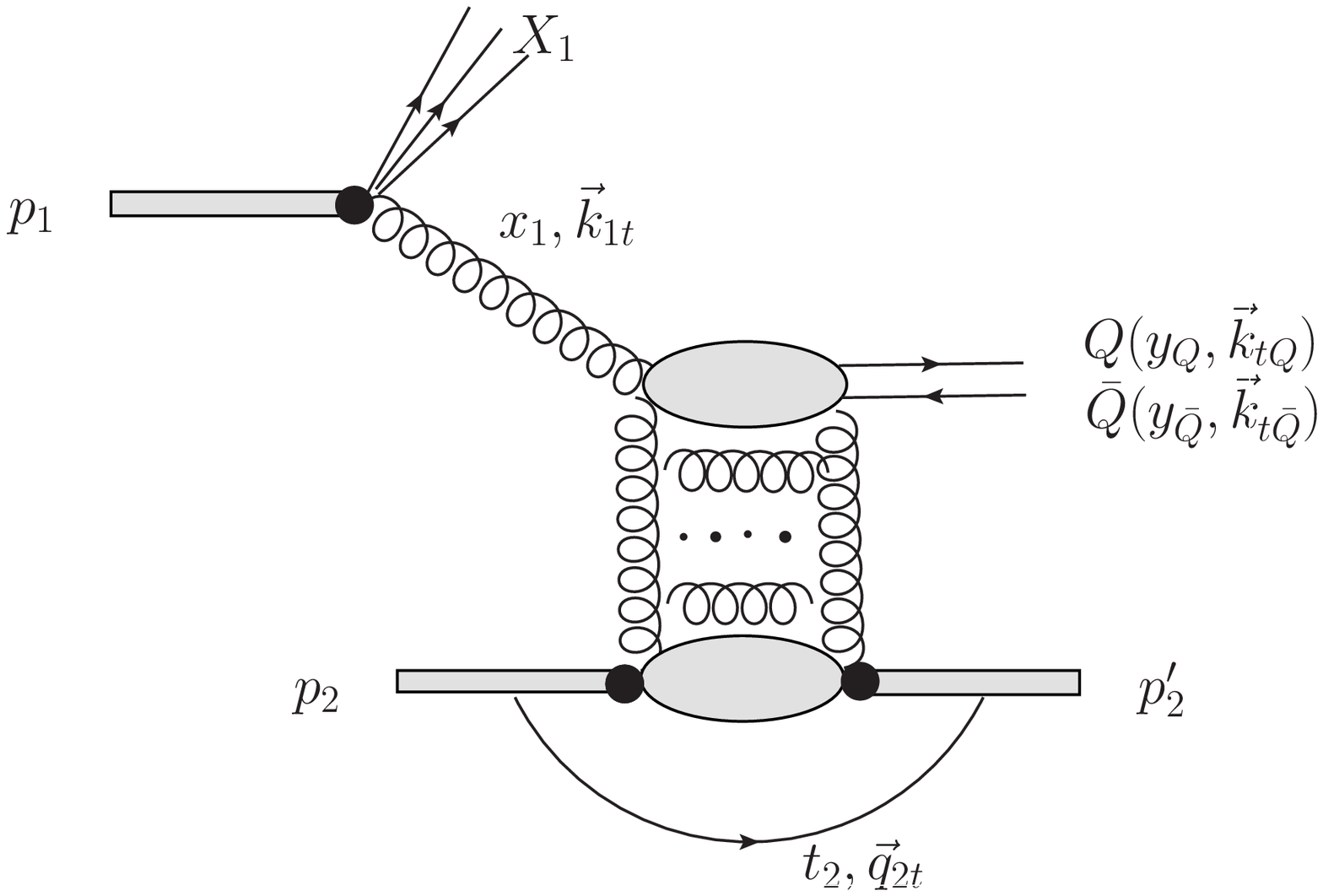}
   \caption{
\small The mechanism of gluon dissociation into $Q \bar Q$ via 
exchange of gluonic ladder in proton-proton collisions.
}
 \label{fig:sd_diagrams}
\end{figure}

The generic mechanism of the reaction is shown in Fig.\ref{fig:sd_diagrams}. 
In this approach $Q \bar Q$ pairs are produced in the Pomeron
fragmentation region, close to the rapidity gap, whereas gluon-fusion 
populate a large part of the phase space taken up by the diffractively 
produced system and will generally give a tiny contribution in the
Pomeron fragmentation region, unless there are a lot of hard gluons in 
the Pomeron.

The mechanism discussed here was considered 
previously in \cite{Alves:1996ue,Yuan:1998ht} in an approximation
in which gluon transverse momenta in the Pomeron are integrated out.
Our results appear to be different from those presented in 
\cite{Alves:1996ue,Yuan:1998ht}.
Somewhat related, but different, microscopic mechanisms were 
discussed in \cite{Kopeliovich:2007vs}).

In our recent paper \cite{LSS2013} we have presented the amplitude for 
the $g p \to Q \bar Q p$ (sub)process and calculateed
integrated and differential cross section for the $p p \to Q \bar Q p X$
single-diffractive processes at the LHC.
Here we show only some selected results.

\section{Sketch of the formalism}

It is shown in our recent paper \cite{LSS2013} how to obtain 
the amplitude for the $g p \to Q \bar Q p$ process. 
The differential parton-level cross section can be written as: 
\begin{eqnarray}
16 \pi {d\hat \sigma(g N \to Q\bar Q N;\hat{s}) \over d\bDelta^2}\Big|_{\bDelta^2=0} = {1 \over 2 \cdot (N_c^2-1)} \cdot
\sum_{\lambda_g,\lambda,\bar\lambda,a} \Big| {\cal{A}}_D(g_{\lambda_g}^a N \to Q_\lambda \bar Q_{\bar \lambda} N) \Big|^2 \, dz 
{d^2\bk \over (2 \pi)^2} \, , 
\nonumber \\
\end{eqnarray}
and the final multi-dimensional cross section reads:
\begin{eqnarray}
{d \hat \sigma (gN \to Q \bar Q N;\hat{s}) \over dz d^2\bk d\bDelta^2}\Big|_{\bDelta^2=0} = {\pi  \over 4 \, N_c^2 (N_c^2-1)^2} \,\alpha_S 
\Big \{ [z^2 + (1-z)^2] \bPhi_1^2 + m_Q^2 \Phi_0^2 \Big\}  \, .
\label{eq:parton-level}
\end{eqnarray}
The auxiliary functions $\Phi_0$ and $\bPhi_1$ are defined in 
Ref.\cite{LSS2013}.
%
and 
%
Finally we calculate the spectrum of quarks in 
the $pp$-collision. Starting from the diffractive 
$g p \to Q \bar Q p$ cross section
\begin{eqnarray}
 {d \hat \sigma (gN \to Q \bar Q N;\hat{s}) \over dz d^2\bk d\bDelta^2}\Big|_{\bDelta^2=0} = \hat f_{Q \bar Q}(z,\bk;\hat s) \, 
\end{eqnarray}
we can obtain the corresponding cross section for $pp$-collisions in 
the collinear approximation for the incoming gluon as:
\begin{eqnarray}
{d \sigma (pp \to X Q\bar Q + p;s) \over dx_Q d^2\bk d\bDelta^2}\Big|_{\bDelta^2=0} &=&
\int dx dz \, \delta(x_Q - xz) \, g(x,\bar Q^2) \,  \hat f_{Q \bar Q}(z,\bk;xs) \, 
\; ,
\nonumber \\
&=& \int_{x_Q}^1  {dx \over x} g(x,\bar Q^2) \,  \hat f_{Q \bar Q} \Big( {x_Q \over x}, \bk; x s \Big) \, .
\end{eqnarray}

We can also calculate the  fully differential distribution in 
$x_Q, x_{\bar Q} = x - x_Q$, rapidities etc.
\begin{eqnarray}
{d \sigma (pp \to X Q\bar Q + p;s) \over dx_Q dx_{\bar Q} d^2\bk d\bDelta^2}\Big|_{\bDelta^2=0} &=&
{1 \over x_Q + x_{\bar Q}}  \, g(x_Q + x_{\bar Q},\bar Q^2) \,  \hat f_{Q \bar Q} \Big( {x_Q \over x_Q + x_{\bar Q}},\bk;xs \Big) \, .
\end{eqnarray}

\section{Results}
\label{section:Results}

In our numerical calculations, we shall use three different 
UGDFs from the literature.
One of them from Ref.\cite{IN2002} (labelled Ivanov-Nikolaev) is a fit
to HERA structure functions data.
The other two, from Ref.\cite{KS2005} (labelled Kutak-Stasto) are
obtained by solving a BFKL equation accounting for subleading terms. 
One of the latter UGDFs also accounts for a 
nonlinear Balitsky-Kovchegov type term in the evolution equation. 
Both UGDF sets give a reasonable description of deep inelastic
structure functions at small $x$.

For the argument of running coupling constant we take: 
$\mu_r^2 = M_{Q \bar Q}^2$ for $g \to Q \bar Q$ splitting and
$\mu^2 = \max(\kappa^2, \bk^2+m_Q^2)$ for the $t$-channel coupling of gluons
to heavy quarks. 
For the quark masses, we take $m_c = 1.5 \, \mathrm{GeV}$ and
$m_b = 4.75 \, \mathrm{GeV}$

\begin{figure}[!h]
\includegraphics[width=7cm]{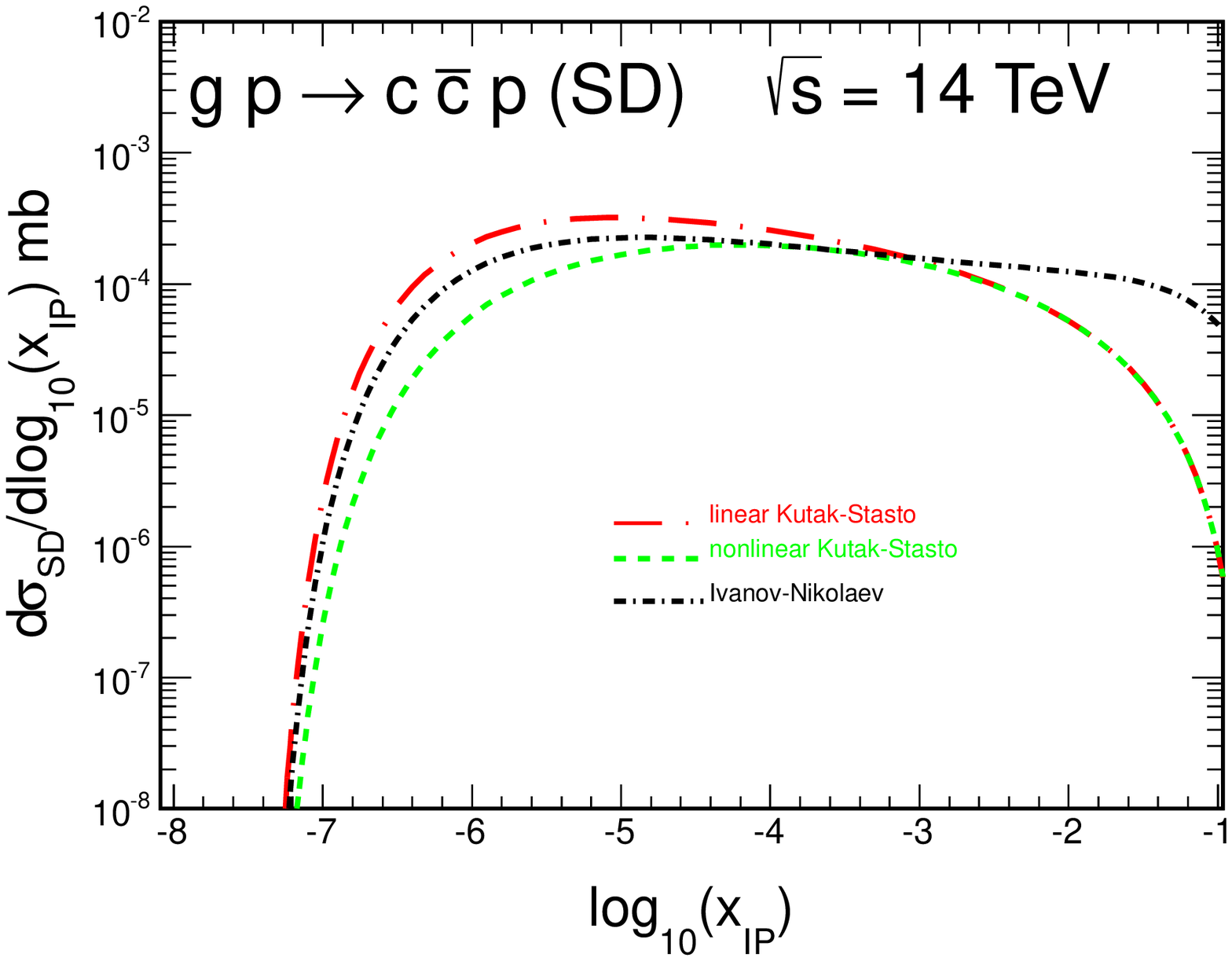}
\includegraphics[width=7cm]{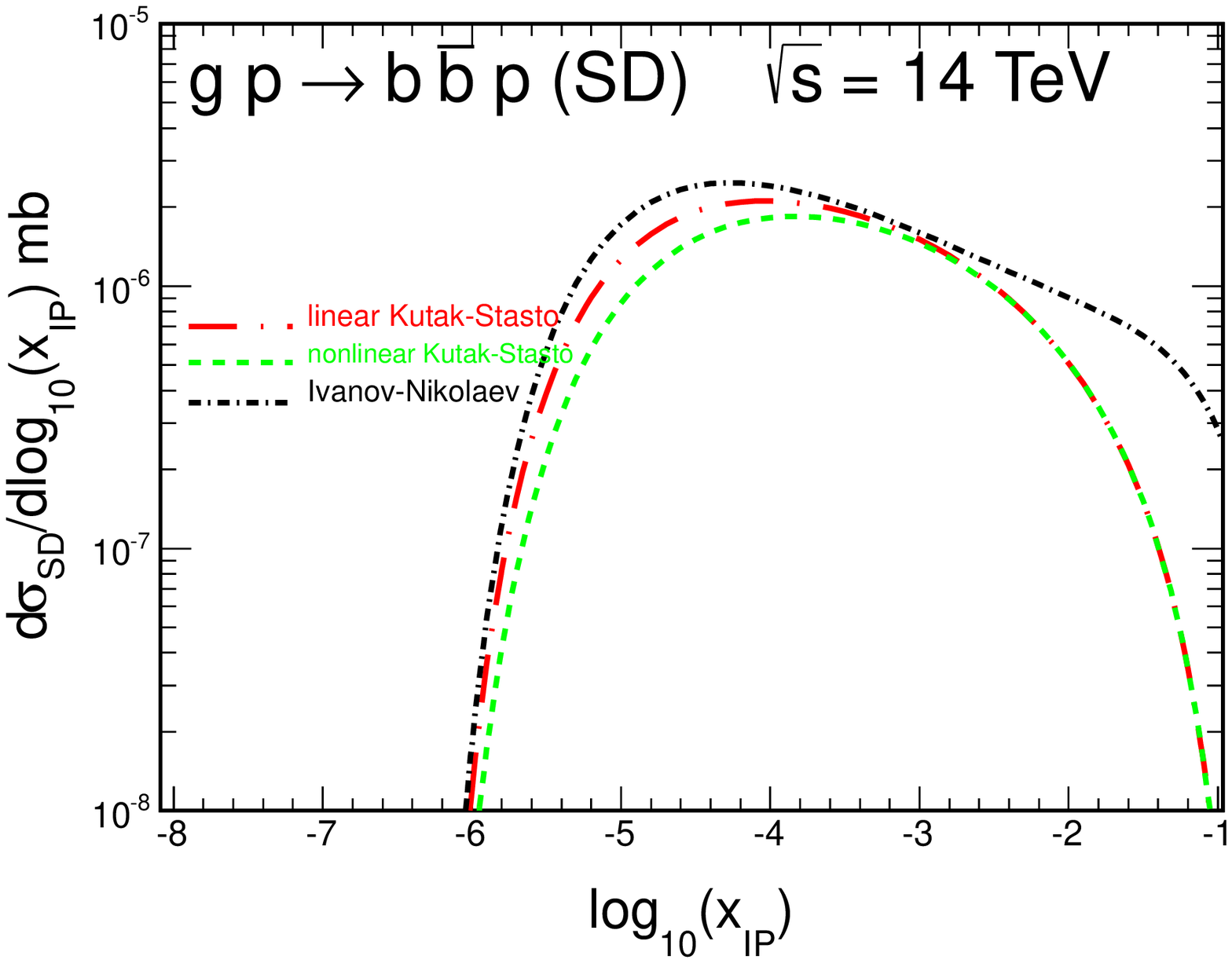}
   \caption{
\small Distribution in log$_{10}(x_{\Pom})$ for $c \bar c$ (left) and 
$b \bar b$ (right) produced in a single diffractive process for 
center of mass energy $\sqrt{s}$ = 14 TeV for the Ivanov-Nikolaev (solid),
linear Kutak-Sta\'sto (dashed) and nonlinear Kutak-Sta\'sto (dotted) UGDFs. 
Absorptive effects have been included by multiplying by
gap survival factor.
}
\label{fig:dsig_dxi2}
\end{figure}

Now differential distributions will be discused.
Here, absorption corrections are included in a rough manner, by
multiplying the cross section by a gap survival factor $S_G$ = 0.05
\cite{KMR_eikonal,Maor}. 
A more subtle treatment, which would include the dependence of
absorption effects on kinematical variables must be developed in the future.

Let us start with distributions in $x_{\Pom}$ -- the fractional longitudinal
momentum loss of proton. Notice that $\log(1/x_{\Pom})$ is proportional to
the size of the rapidity gap. The cross section drops sharply at 
$x_{\Pom} \lsim 10^{-7}$ for charm quarks and $x_{\Pom} \lsim 10^{-6}$ for 
bottom quarks. This is related to the fact that with increasing gap size
we are asking for harder partons in the dissociating
proton. The gluon distribution however drops sharply at large $x$.

In the Ingelman-Schlein model, the gap size-dependence
is described in terms of a universal flux of Pomerons.  
In our microscopic model the $x_{\Pom}$--dependence is driven by the 
dependence of the unintegrated gluon distribution on 
$x_{\mathrm{eff}}=x_{\Pom}$.

\begin{figure}[!h]
\includegraphics[width=7.5cm]{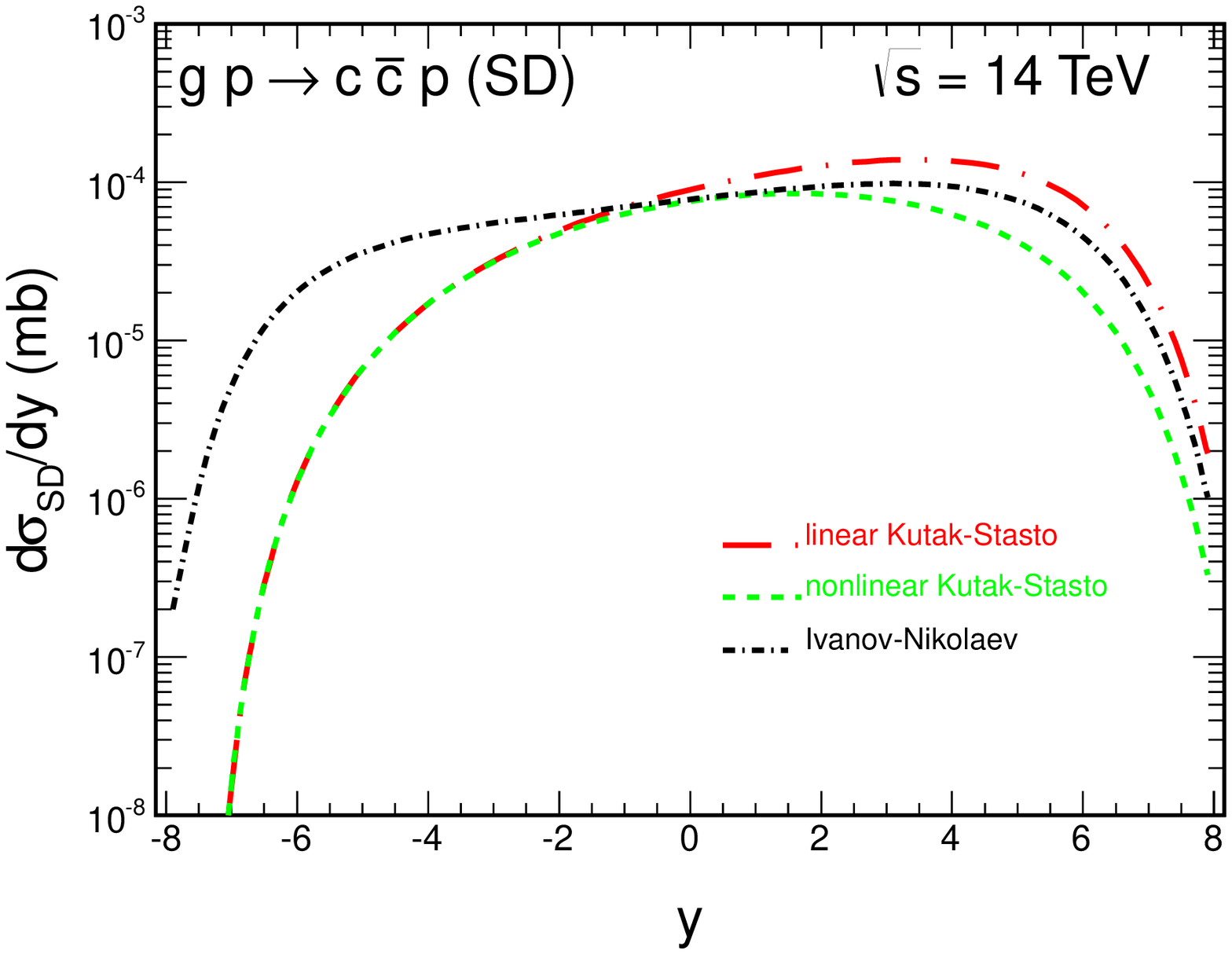}
\includegraphics[width=7.5cm]{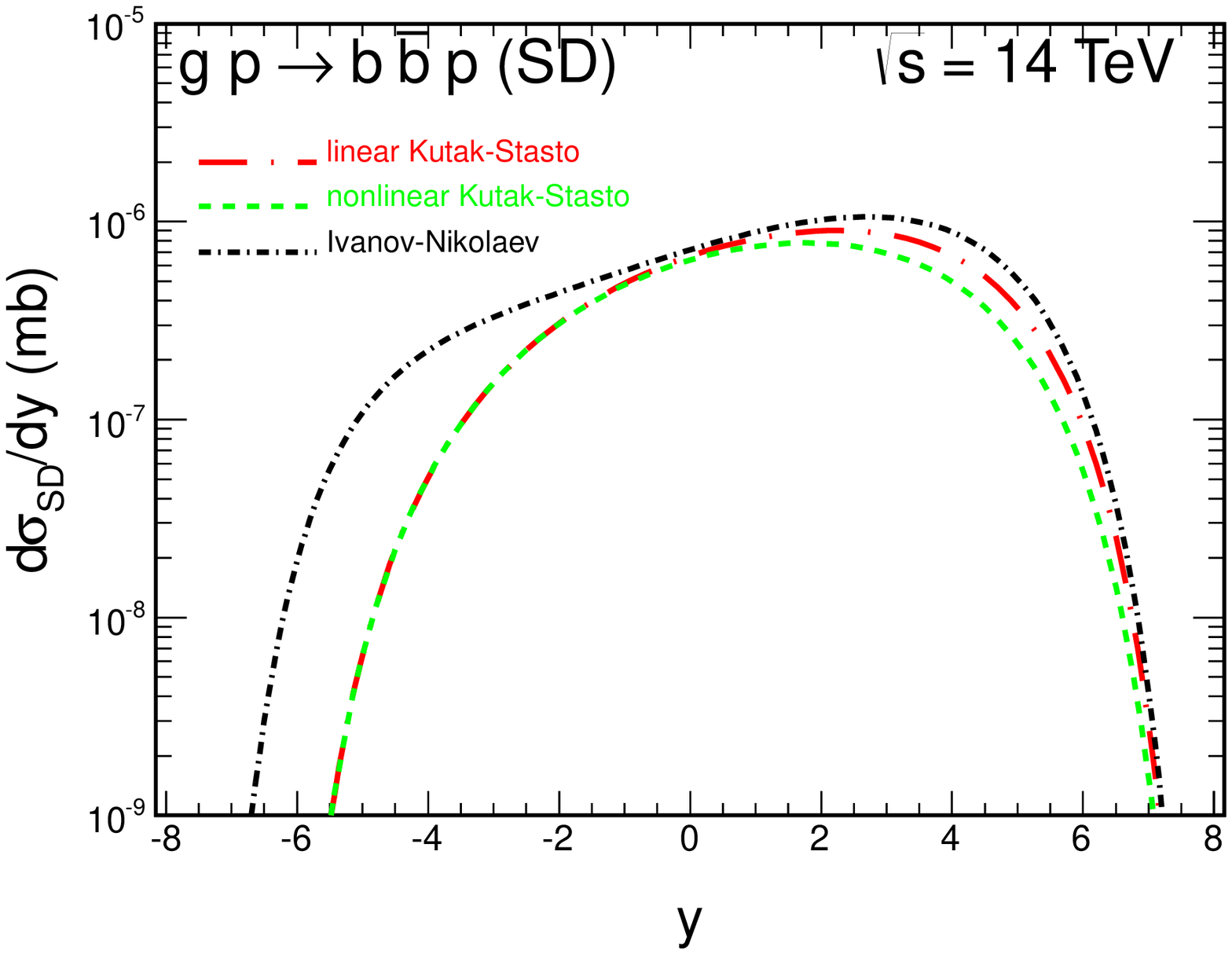}
   \caption{
\small Distribution in rapidity of $c$ ($\bar c$) (left) and $b$ ($\bar
b$) (right) produced in a single diffractive process for center of mass 
energy $\sqrt{s}$ = 14 TeV.
Absorptive effects have been included by multiplying by
gap survival factor.
}
\label{fig:dsig_dy}
\end{figure}

In Fig.\ref{fig:dsig_dy} we present the rapidity distribution of charm (left
panel) and bottom (right panel) quarks/antiquarks from diagram (b)
in Fig.\ref{fig:sd_diagrams}. 
At large rapidities the cross section with the Ivanov-Nikolaev UGDF
is much larger than that with the nonlinear Kutak-Sta\'sto UGDF. 
This is partially due to nonlinear effects included in the latter 
distributions. The nonlinear effects appear at $x_\Pom <$ 10$^{-4}$.

\begin{figure}[!h]
\includegraphics[width=7.5cm]{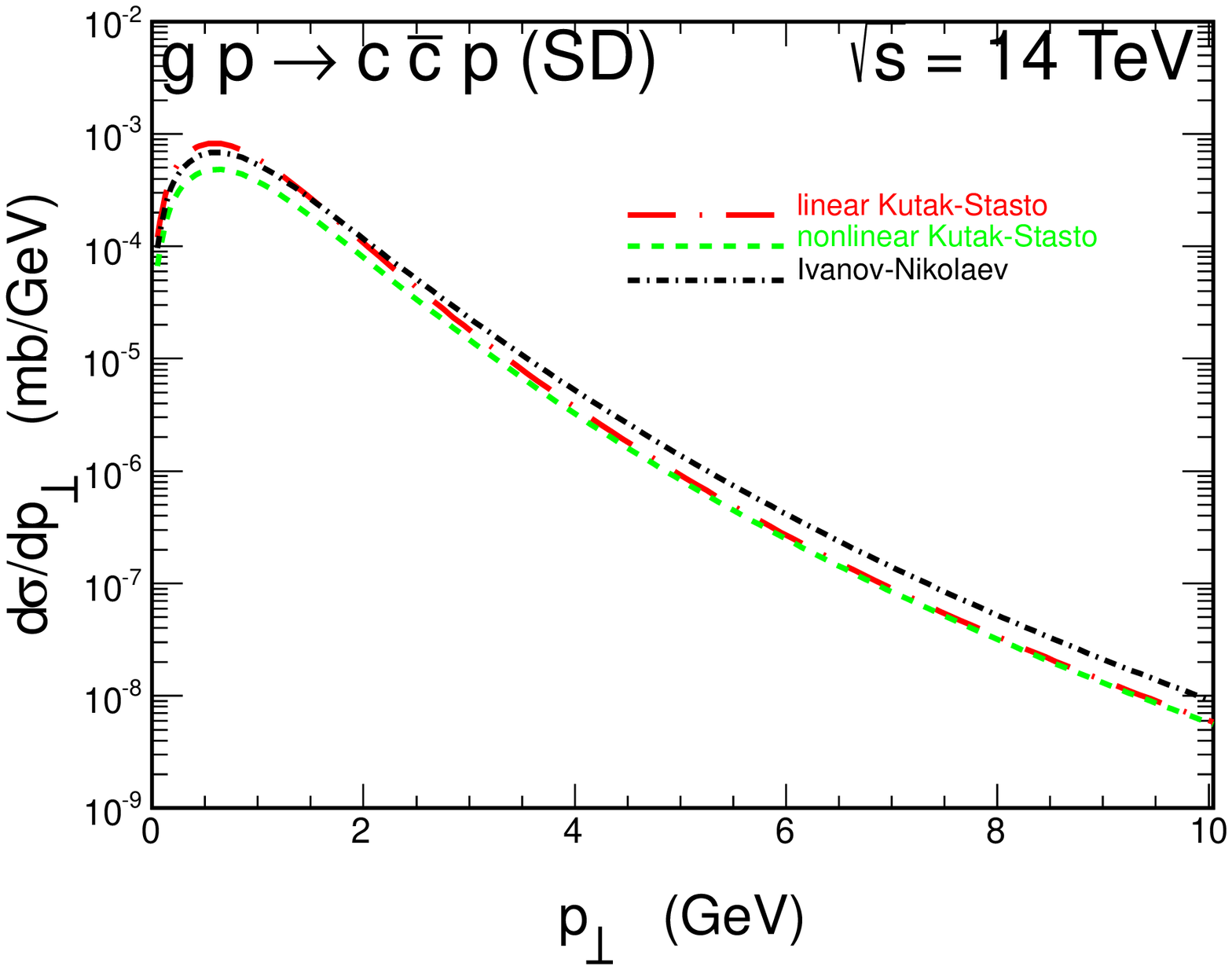}
\includegraphics[width=7.5cm]{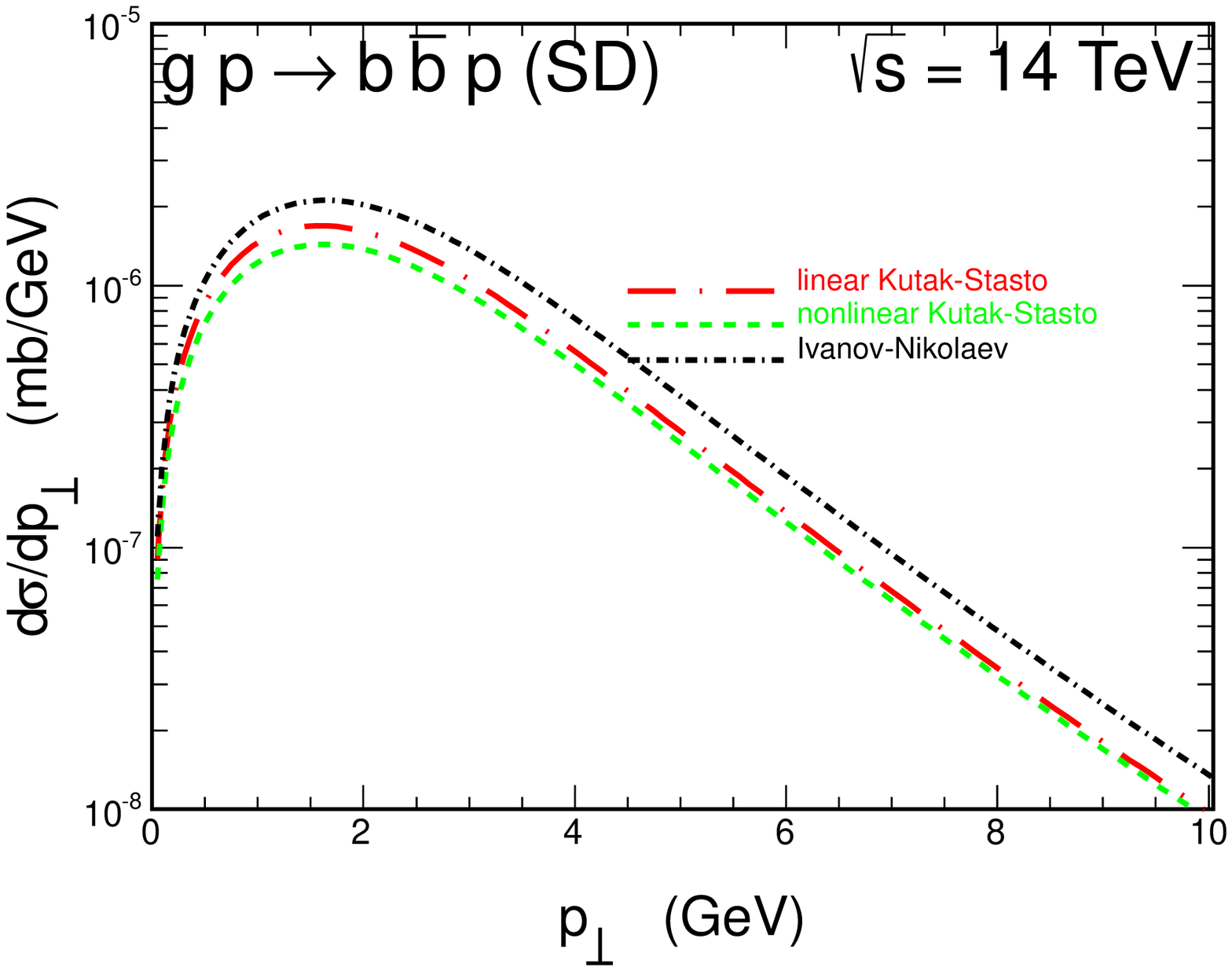}
   \caption{
\small Distribution in transverse momentum of $c$ ($\bar c$) (left) 
and $b$ ($\bar b$) (right) produced in a single diffractive process 
for center-of-mass energy $\sqrt{s}$ = 14 TeV.
Absorptive effects have been included by multiplying by
gap survival factor.
}
\label{fig:dsig_dpt}
\end{figure}

In Fig.\ref{fig:dsig_dpt} we show transverse momentum distributions of 
charm (left panel) and bottom (right panel) quarks/antiquarks from
one single-diffractive mechanism. The spread in transverse momentum here
is somewhat smaller than in the Ingelman-Schlein model calculations of 
Ref.\cite{LMS2011}.

Now we wish to study dependence of the ratio of cross sections
for $b \bar b$ and $c \bar c$ production as a function of some
kinematical variables. Such ratios, to a good approximation, should be
independent of absorption effects. 

\begin{figure}[!h]
\begin{center}
\includegraphics[width=8cm]{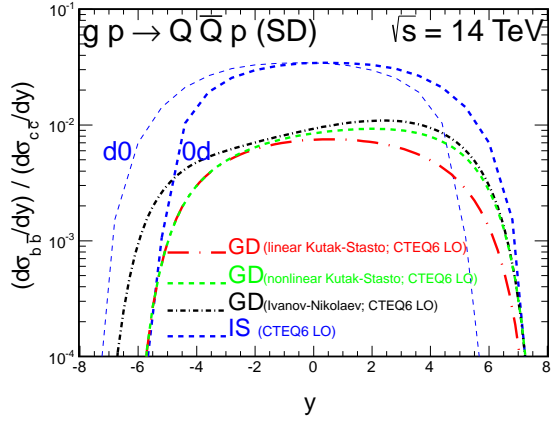}
\end{center}
   \caption{
\small The ratio of the $b \bar b$ to $c \bar c$ distributions in 
quark (antiquark) rapidity.
}
\label{fig:dsig_dy_ratio}
\end{figure}

In Fig.\ref{fig:dsig_dy_ratio} we show the ratio as a function of 
quark rapidity. The ratio for the Ingelman-Schlein model is somewhat 
larger than that for the gluon-dissociation approach.

\begin{figure}[!h]
\begin{center}
\includegraphics[width=8cm]{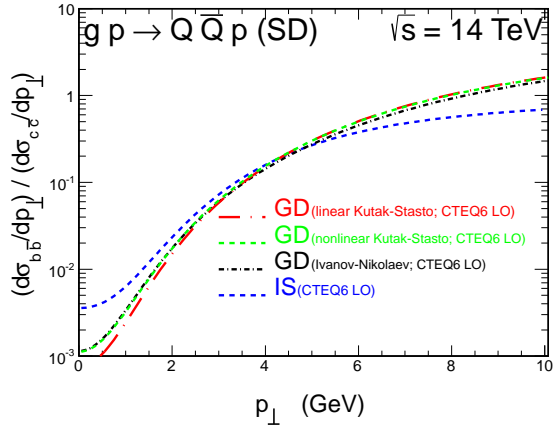}
\end{center}
   \caption{
\small The ratio of the $b \bar b$ to $c \bar c$ distributions in 
quark (antiquark) transverse momentum.
}
\label{fig:dsig_dpt_ratio}
\end{figure}

The charm-to-bottom ratio as a function of transverse momentum of 
the (anti)quark is shown in Fig.\ref{fig:dsig_dpt_ratio}. 
The ratio increases as a function of quark transverse momentum. The
character of the function is in principle similar for both approaches.

\section{Conclusions}

Recently we have derived forward amplitudes for the 
$g p \to Q \bar Q p$ subprocess both in the impact parameter and
momentum space representation in the forward scattering approximation.
The amplitude for the off-forward directions within the diffraction cone
was extrapolated by assuming exponential dependence known from other 
diffractive processes.
The forward amplitude for the $g p \to Q \bar Q p$ subprocess has been 
obtained in terms of unintegrated gluon distribution of the target proton. 

The formulae have been used to calculate cross section for the single 
scattering process $p p \to Q \bar Q p X$ as a convolution of the
collinear gluon distributions in the proton and the elementary 
$g p \to Q \bar Q p$ cross section both for charm and bottom production.
When applied to the hadronic collisions, this approach allows one to predict 
heavy quark production ``close to the rapidity gap''. 

We have presented here some results for the rapidity and transverse momentum
distribution of quarks (antiquarks)
at the nominal LHC energy $\sqrt{s}$ = 14 TeV. The cross section
for charm quarks is two orders of magnitude larger than that for bottom 
quarks, as expected from the $m_Q^{-4}$ scaling of the partonic subprocess.

We have calculated also ratio of the cross section for $b \bar b$
and $c \bar c$ as a function of several kinematical variables.
The ratio is fairly smooth in (anti)quark rapidity and strongly depends
on (anti)quark transverse momentum. 


A measurement of the single diffractive production would be possible
e.g. at ATLAS detector by using so-called ALFA detectors for measuring 
forward protons and their fractional energy loss and the main central 
detector for the measurement of $D$ or $B$ mesons. CMS+TOTEM is another 
option.


\end{document}